\documentclass{tmex20}

\def\Journal#1#2#3#4{{#1} {\bf #2}, #3 (#4)}

\def\PRL{\em Phys. Rev. Lett.}
\def\PRD{{\em Phys. Rev.} D}

\def\be{\begin{equation}}
\def\ee{\end{equation}}
\def\bea{\begin{eqnarray}}
\def\eea{\end{eqnarray}}

\newcommand{\Photo}{}

\begin{document}
\vspace*{4cm}
\title{Search for low-mass dark matter with the DAMIC experiment}

\author{Mariangela Settimo for the DAMIC and DAMIC-M collaboration }

\address{SUBATECH, IMT-Atlantique, CNRS/IN2P3, Universit\'e de Nantes, Nantes (France)}

\maketitle\abstracts{
The DAMIC (Dark Matter in CCDs) experiment searches for the interactions of dark matter particles with the  nuclei and the electrons in the silicon bulk of thick fully depleted charge-coupled devices (CCDs). Because of the low noise and low dark current, DAMIC CCDs are sensitive to the ionization signals expected from low-mass dark matter particles ($< 10$~GeV). A 40-gram target detector has collected data at the SNOLAB underground laboratory since 2017. Recent results from the searches for DM-electron scattering and hidden-photon absorption will be summarized and the status of WIMPs-nucleon search reported. A new detector --  DAMIC-M (DAMIC at Modane) -- with a mass-size of 1 kg and improved CCD readout is under design and will be installed at the underground laboratory of Modane, in France. The current status of DAMIC-M  and the near future plans will be presented. }

\section{Introduction}

Identifying the nature of dark matter (DM)  is one of the biggest challenges in astroparticle physics and cosmology today. 
Despite the impressive improvements in sensitivity by noble liquid detectors, no positive signals have been observed so far by experiments looking WIMPs with masses of tens or hundreds GeV. 
Other viable candidates are thus receiving increased attention, in particular sub-GeV WIMPs or DM particles from a hidden-sector~\cite{subGeV}, which couple weakly with ordinary matter through, for example, mixing of a hidden-photon with an ordinary photon. These particles may interact with electrons with sufficient energy transfer to be detectable. Moreover, eV-mass hidden photons can be DM candidates them-self and they can be absorbed by electrons in the sensitive material. 

The DAMIC (DArk Matter In CCD) experiment~\cite{DAMIC} employs the bulk silicon of scientific-grade charge-coupled devices (CCDs) as target for the interactions of dark matter (DM) particles. 
Thanks to the low pixel readout noise and low leakage current DAMIC extremely sensitive to the faint ionization signals expected from the interaction of DM with nuclei or electrons in the silicon bulk. The low mass of the silicon nucleus provides good sensitivity to WIMPs with masses in the range 1-10 GeV/c$^{2}$, while the small band gap of silicon provides sensitivity to DM-electron interactions that deposit as little as 1.1 eV in the target. 

An R\&D program has been conducted since 2013 to prove the performance of the CCD~\cite{DAMIC,compton,ionisation} and provide measurements of the background contamination~\cite{background}. Limits on WIMP-nucleon cross-section\cite{WIMPlimits},  hidden photon absorption~\cite{HiddenPhotons} and DM-electron scattering~\cite{DMelectron-scattering} with only a few grams of target mass.
A detector with a total mass of 40~g is currently installed at the SNOLAB~\cite{Snolab} underground laboratory (a mine 2000 m underground in Canada) and has been collecting data for calibration, radioactivity studies and for DM searches. This proceeding will summarise the latest published results and updates on the analysis ongoing with the data from the latest prototype. 

\section{The DAMIC experimental setup}

In DAMIC, the sensitive detector is the silicon bulk of high-resistivity fully depleted CCDs. Each device is a 16 Mpixels CCD, with a pixel size of (15$\times$15)$\rm{\mu m^2}$, a thickness of 675~$\rm{\mu m}$ and a mass of 5.9~g.

\begin{figure}
\centering
\includegraphics[width=0.4\linewidth]{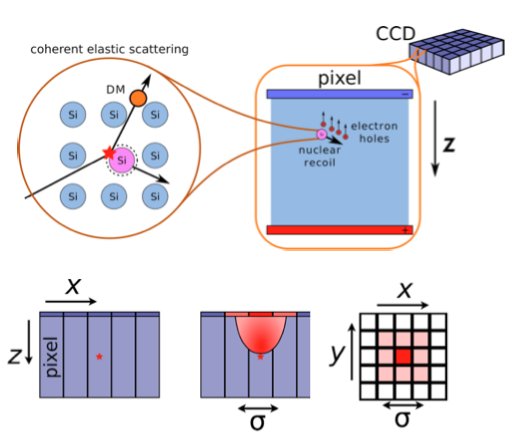}
\hspace{10mm}
\includegraphics[width=0.35\linewidth]{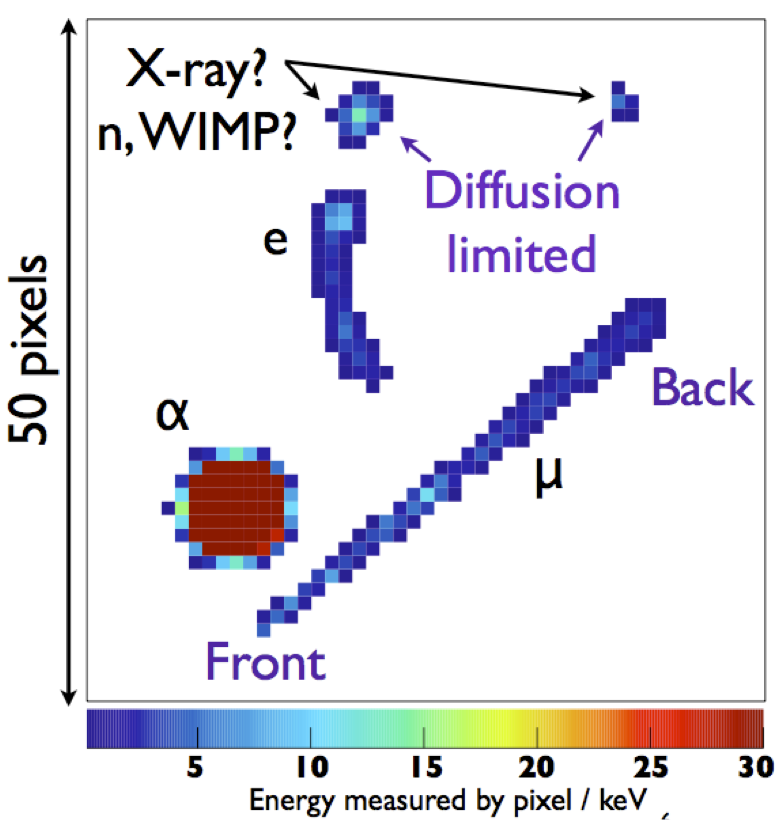}
\caption[]{Left: Representation of the WIMP-nucleon scattering and of the charge diffusion by a point-like ionization event in the CCD bulk. The x-y coordinates give the position in the CCD whereas the lateral spread positively correlates to the depth of the energy deposit. The diffusion model has been tested with data from radioactive sources and cosmogenic muons. Right: example of  detected tracks from different particle types.}
\label{fig:principle}
\end{figure}

When a DM particle scatter off the silicon nuclei or electrons (Fig.~\ref{fig:principle} left), the ionization charge is drifted along the direction of the electric field (z axis) and collected on the pixel array (x-y plane). Because of thermal motion, the charge diffuses transversely with a spatial variance ($\sigma_{xy}$) that is proportional to the transit time (i.e., the depth of the interaction point). This allows for the reconstruction in three dimensions of the position of the energy deposit in the bulk of the device, and the identification of particle types based on the cluster pattern (Fig.~\ref{fig:principle} right). The diffusion parameters are calibrated using  tracks of traversing muons~\cite{WIMPlimits} and X-ray events interact near the surfaces on the back and the front of the CCD~\cite{DAMIC2014}.

The exquisite spatial resolution and the 3D reconstruction allow for the rejection of background events produced by low energy gammas and electrons on the surface of the CCD and for the characterization of the radioactive background on the surface and in the bulk of the CCD~\cite{background}. In particular, DAMIC has unique chance for the  measurement of the U/Th contamination and of Pb-210 and Si-32 in the bulk of the detector. Measurements are based on the identification of the respective decay chains by looking at the spatial coincidence of betas and gammas. As an example, the isotope Si-32 in the bulk of silicon is produced by cosmogenic activation and $\beta$-decays with a life time of about 150 years in P-32 with subsequently  $\beta$-becay with a life-time of about 14 days. The resulting $\beta$ spectrum extends to low energies becoming an irreducible background for all silicon-based experiments (e.g., DAMIC, SuperCDMS~\cite{SuperCDMS}). An update of these measurements with the new DAMIC CCDs has been presented in~\cite{Matalon}. A contamination of $133.3 \pm 27.8{\rm \mu Bq/kg}$, $83.1 \pm 11.8 {\rm nBq/cm^2}$ have been measured for Si-32 and Pb-210 respectively. For the U-238 and Th-232 chains we search for a $\alpha-\beta$ and a $two-\alpha$ coincidences but the absence of these characteristic signatures constraints the activity to less than 0.53~/kg/day and 0.35~/kg/day, respectively. Compared to the CCD used in the previous DAMIC prototype of 2015~\cite{background} we thus observe a significant improvement in the Si-32 and Pb-210 due to a better material selection and CCD production. 

\begin{figure}[tb]
\centering 
\includegraphics[width=0.6\linewidth]{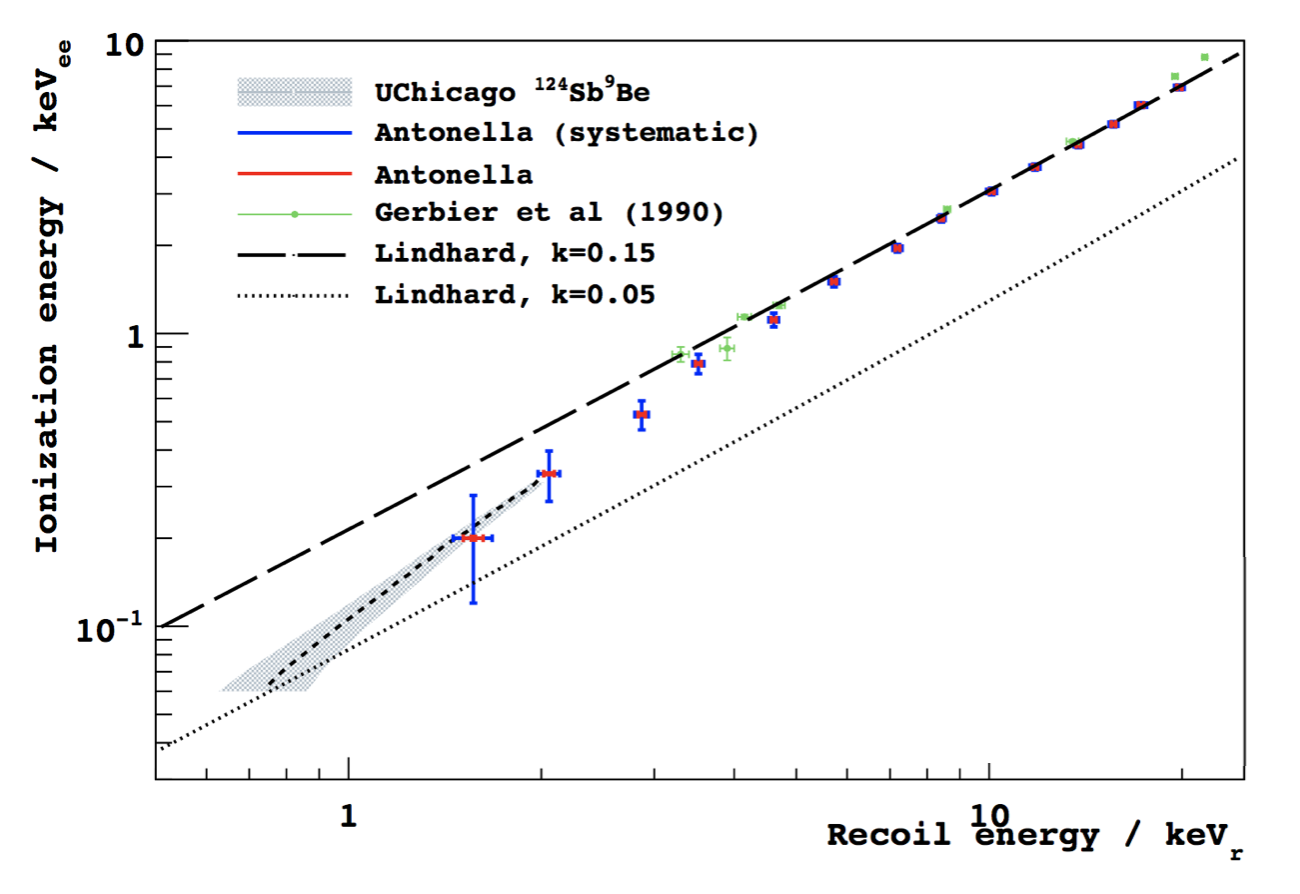}
\includegraphics[width=0.34\linewidth]{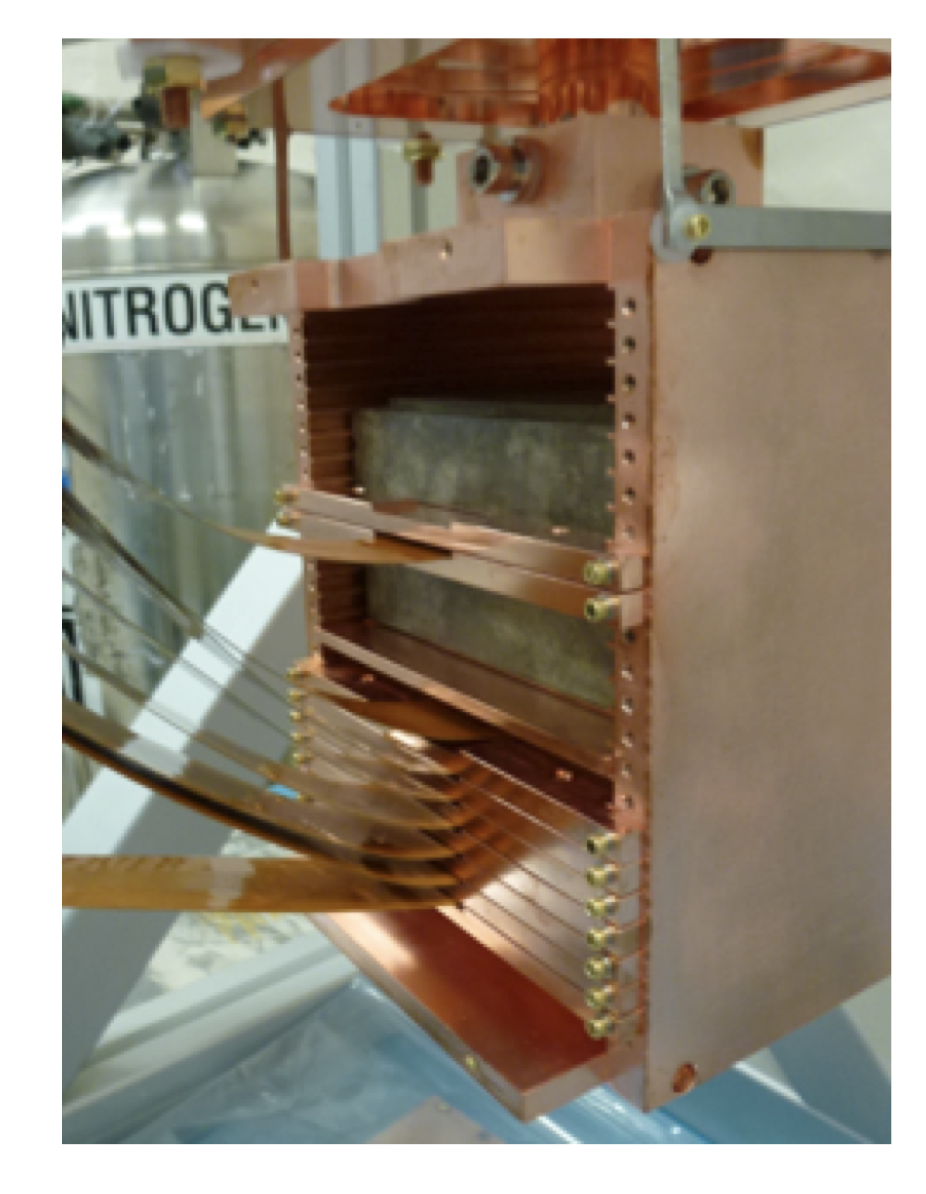}
\caption[]{Ionisation signal ($E_e$) vs recoil energy ($E_r$) in silicon. Results are from  $^{124}$Sb-$^{9}$Be photo-neutron source experiment\cite{ionisation} and a pulsed fast-neutron beam\cite{antonella} are shown as short dashed line (with a gray band representing the 1$\sigma$ uncertainty)  and red markers, respectively. The extrapolations of the theoretical Lindhard model is shown for reference.}
\label{fig:quenching}
\end{figure}

The response of DAMIC CCDs to ionizing radiation has been measured with optical photons~\cite{DAMIC} for ionization signals down to 10 electrons, and with mono-energetic X and  $\gamma$-ray sources for energies in the range 0.5-60 keV$_{ee}$~\footnote{eV$_{ee}$ is the electron-equivalent energy scale relative to the ionization produced by recoiling electrons.}. In the case of recoiling nuclei the produced ionization signal (number of electron-hole pairs) is smaller than the one from recoiling electron of the same kinetic energy. The ionization efficiency in the energy range covered by DAMIC has been measured comparing the observed and predicted nuclear-recoil energy spectra in a CCD from a low-energy $^{124}$Sb-$^{9}$Be photo-neutron source~\cite{ionisation} and using a pulsed fast-neutron beam with a silicon drift detector~\cite{antonella}. A deviation from the extrapolated values of the theoretical Lindhard model~\cite{Lindhard} has been pointed out at energies below 3 keV (see Fig.~\ref{fig:quenching} left). 

The current detector at Snolab consists of 7 CCDs (one of them sandwiched between two blocks of ancient lead) for a total mass of 40~g (Fig.~\ref{fig:quenching} right).  Each CCD is epoxied onto a silicon backing, together with a flex cable that is wire bonded to the CCD and provides the voltage biases, clocks and video signals required for its operation. These components are supported by a copper frame to complete the CCD module. The modules are installed in a copper box that is cooled to 140 K inside a vacuum chamber. The box is shielded by 18~cm of lead to attenuate external $\gamma$ rays, with the innermost 2-inches made of ancient lead, and by  42~cm-thick polyethylene to moderate and absorb environmental neutrons. Each image is acquired with an exposure of $3 \times 10^4$ and $10^5$ seconds (about 8 and  24 hours) and is followed by a zero-length exposure (``blank'') for noise and detector monitoring.  
Fig.~\ref{fig:analysis} (left) shows the pixel value distribution as observed in blanks images in blue (white noise) and in 8 hours exposure in red (i.e., the convolution of white noise, leakage current and signal). The latter distribution is consistent with the one expected for a  white noise, and is well described by a Gaussian distribution with a standard deviation of $\sim 1.6$~electrons (about 6 eV$_{ee}$) and a leakage current as low as 10$^{-3}$~e$^{-}$/pix/day. 

 A data set corresponding to about  7.6~kg~day exposure, has been recorded for background studies and data characterization and about 13~kg day exposure data are used for dark matter searches. In the calibration and background runs, the charge collected by each pixel is read out individually to maximize spatial resolution. In the second data set, the charge are collected by column segments of 100 pixels in a single readout. The signal over noise is thus improved and the energy threshold reduced. Since December 2018 more runs where taken for studies on leakage current evolution and  the system is currently dedicated to R\&D.

\section{WIMPs-nucleon scattering}
\begin{figure}[bt]
\centering 
\includegraphics[width=0.50\linewidth]{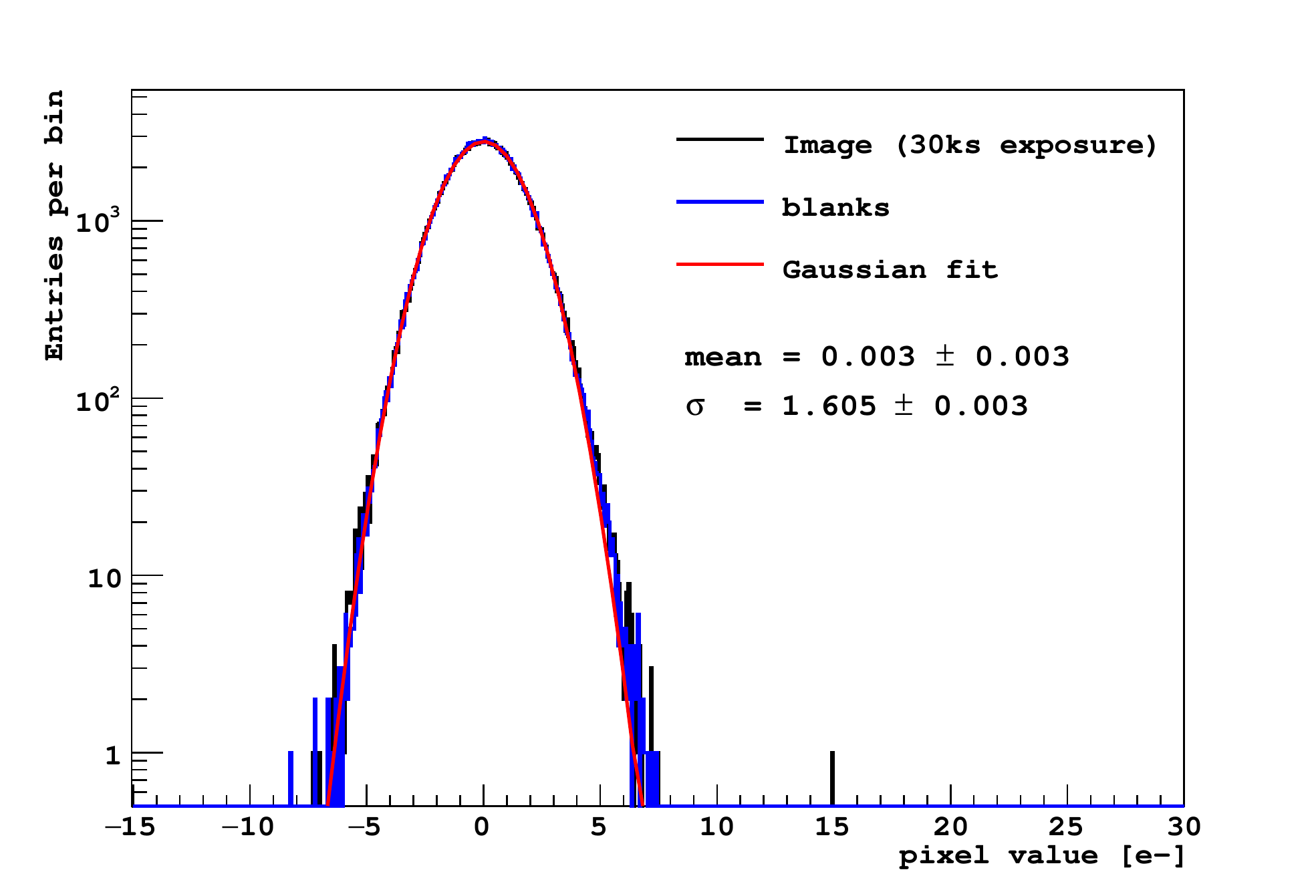}\hfill
\includegraphics[width=0.50\linewidth]{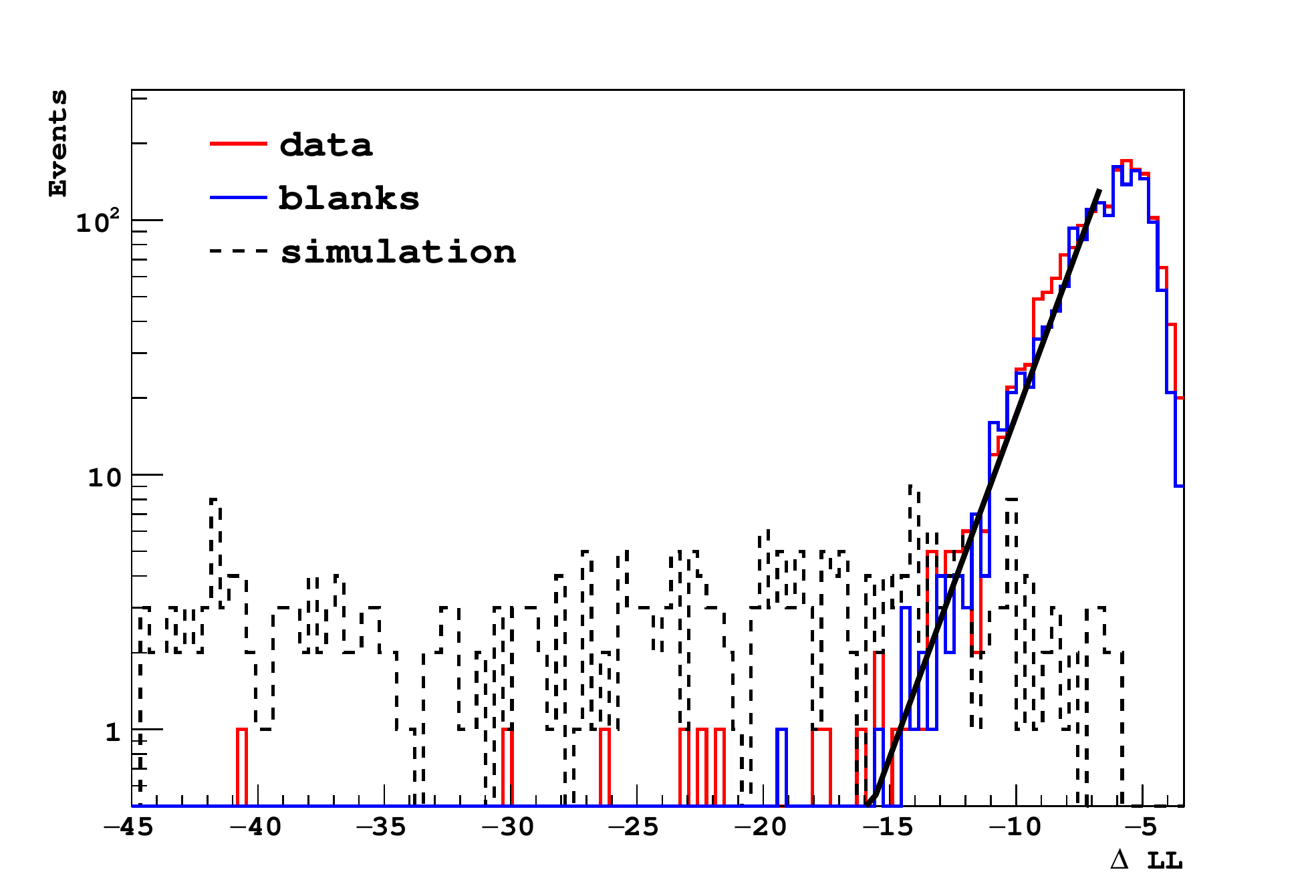}
\caption[]{Left: pixel value distribution in blanks images (white noise) and in 8 hours exposure (white noise, leakage current and signal) when operating at 140~K. 
Right: $\Delta_{LL}$ distributions for all clusters, in the blanks (blue), in simulations (black) and in data (red). }
\label{fig:analysis}
\end{figure}

For WIMPs-nucleon interactions we search for diffuse limited clusters due to the energy deposits in the CCD. 
To select these clusters  we perform a scan of the image with a moving window: For every position of the window, we compute the likelihood $\mathcal{L}_n$ that the pixel values in the window are described by white noise and the likelihood $\mathcal{L}_G$ that the pixel values in the window are described by a Gaussian function plus the white noise. Large (or less negative) values of $\Delta LL = \mathcal{L}_n - \mathcal{L}_G$ are expected from noise events, as shown in Figure~\ref{fig:analysis},  for the blank images (blue), the simulated energy deposits (dashed black) and for real images (red) where events due to noise and energy depositions are both present. 
A selection on the value of $\Delta$LL can be defined to efficiently reject clusters that arise from readout noise.  With the current data an energy threshold of 50 eV$_{ee}$ can be achieved when accepting a leakage from readout noise  of 0.01 events in the entire data set.

In a previous analysis, to select a data set of events with reduced background contamination, surface events are rejected applying a fiducial region cut based on the diffusion parameter $\sigma_{xy}$~\cite{Moriond2018}. The background level obtained in the bulk of the CCD (after surface events rejection)  between 0.5 keV and 14.5 keV has been estimated between as 5 dru\footnote{1 dru (differential rate unit) corresponds to 1 event/kg/day/keV.} (2 dru in the lead sandwiched CCD). 
To avoid loosing exposure because of the fiducialisation, a new approach based on the construction of a background model of radioactive contaminants in the detector has been implemented. A detailed description of the model construction has been presented in~\cite{backgroundTAUP2019}. 
Detailed detector simulations have been developed  using GEANT4~\cite{Geant4}. 
Long-lived isotopes in detector components (for a total of 23 isotopes across 65 detector volumes) have been simulated and the energy and depth of the energy deposits recorded.  
The diffusion model, readout noise, dark current and saturation are taken into account to reproduce clusters as in data. The same reconstruction algorithm is applied in simulations as in data to measure the energy ($E$) and the depth estimator ($\sigma$). 
We then construct a template fit using weighting each component by a parameter that is fixed to the activity of the isotope if it has been measured (typically by ICP-MS,  germanium gamma counting or previous analyses). In the other cases the parameter is left free to float down to zero if only upper bounds on the activity are available or use it unconstrained otherwise. 
The fit is performed in the region above 6~keV. We then use the best fit template to simulate the background in the region of interest (50~eV - 6~keV). From the template fit, we derive important indications on the main sources of background  which are the Pb-210 on various surfaces of the detector, the impurities in the copper material, the cosmogenic activation of silicon bulk and the hydrogen captured on the backside of the dead layer of the CCD. The latter has been confirmed by dedicated measurement performed with secondary ion mass spectroscopy on the back inside of the active region. 

\begin{figure}
\includegraphics[width=0.55\linewidth]{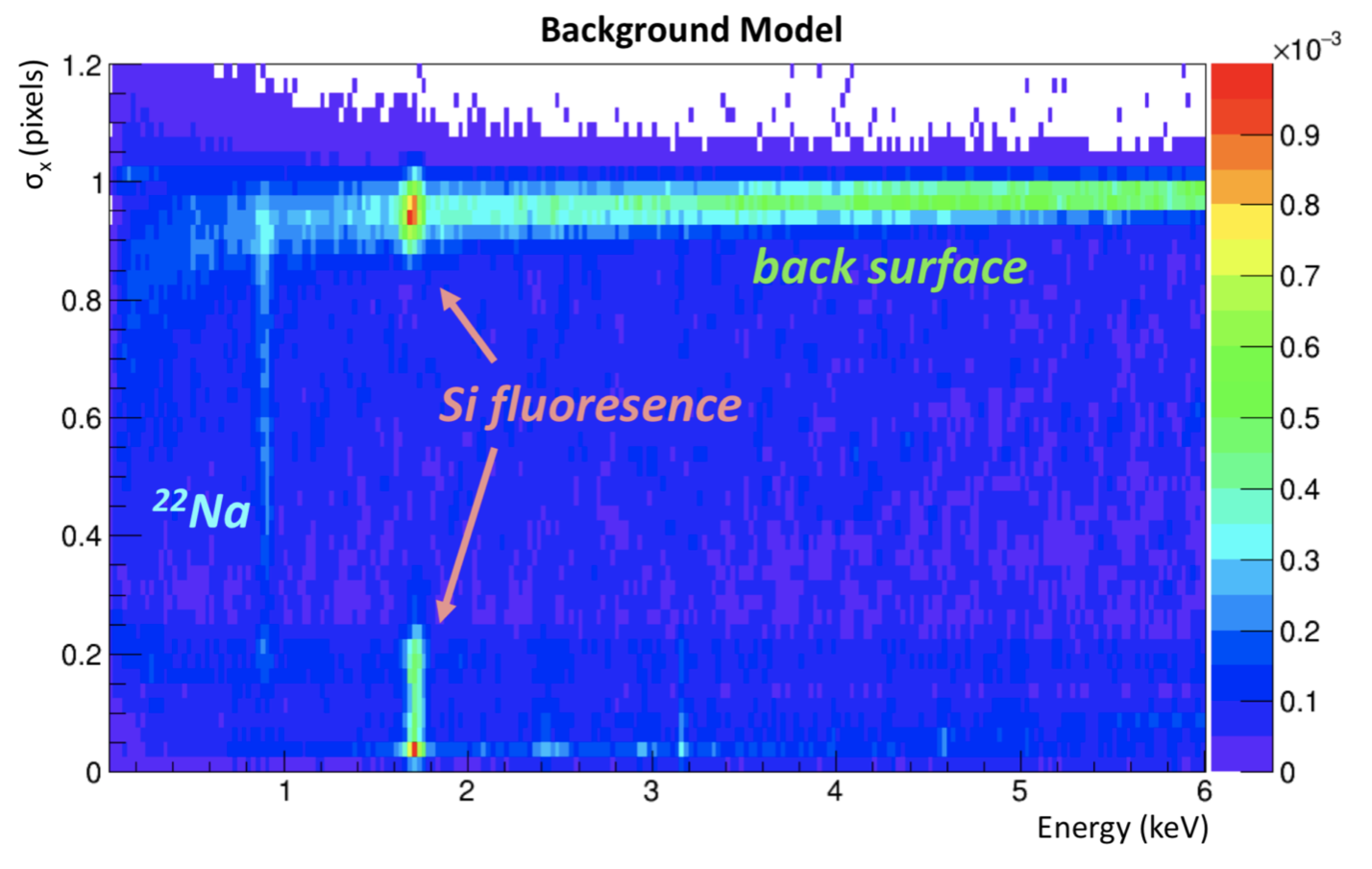}
\includegraphics[width=0.45\linewidth]{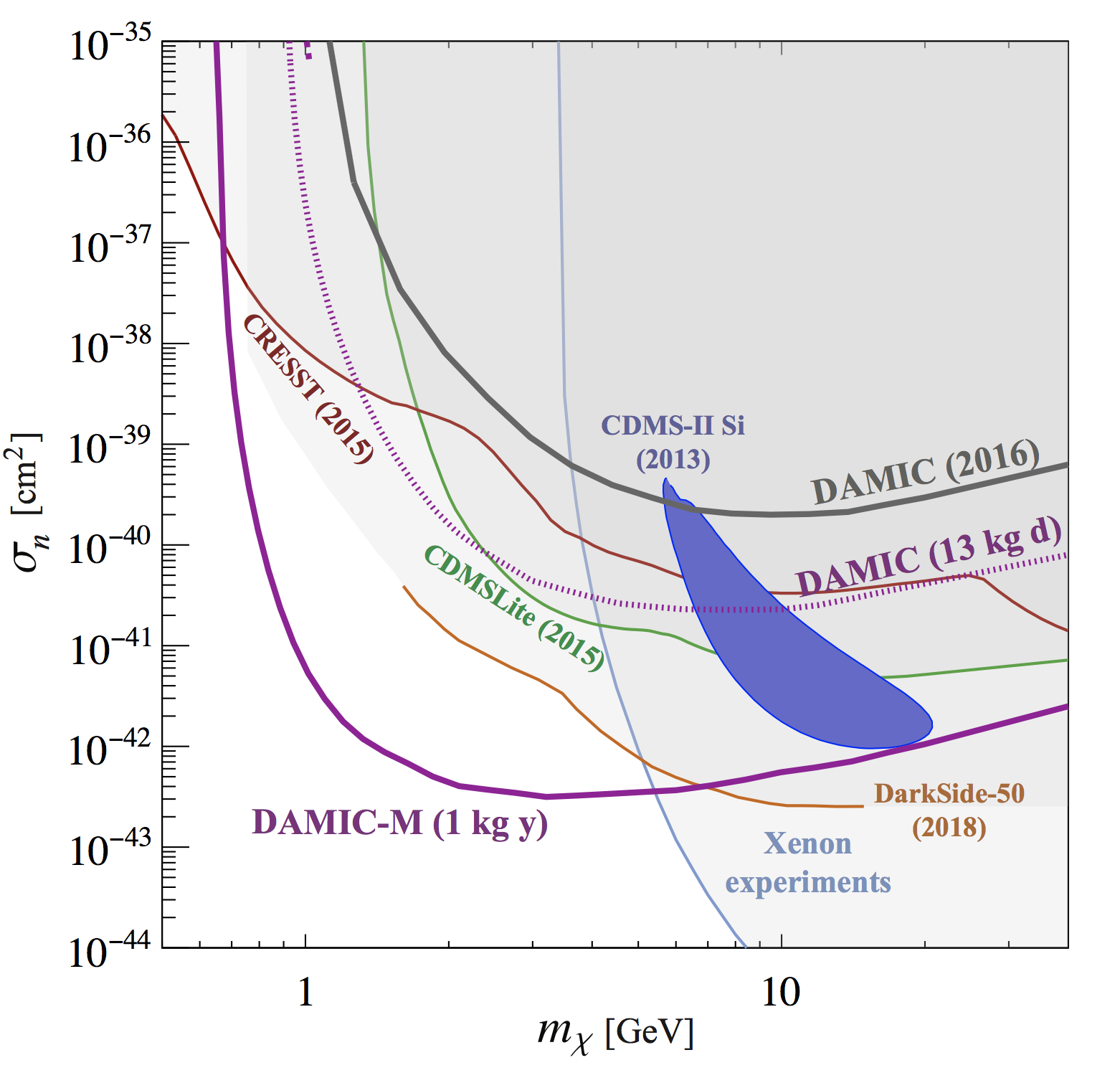}
\caption{Left: Background model in (E, $\sigma_{x}$) phase-space, constructed from the Monte Carlo simulations. Right: Expected sensitivity of DAMIC100 and DAMIC-M for WIMP-nucleon spin-independent scattering. Exclusion limits from other dark matter searches are shown for comparison.}
\label{fig:background}
\end{figure}

Figure~\ref{fig:results} (left) shows the expected sensitivity of DAMIC to the spin-independent WIMPs-nucleus cross-section with a 13~kg~day exposure (pointed). With the current exposure, we can explore a large parameters phase-space of the signal-excess reported by the CMDS collaboration~\cite{CDMS} using the same nuclear target and an energy threshold that is improved by a order of magnitude compared to CDMS-II Si. The WIMP analysis with the background model is in progress and will be finalized soon.

\begin{figure}[tbh]
\includegraphics[width=0.51\linewidth]{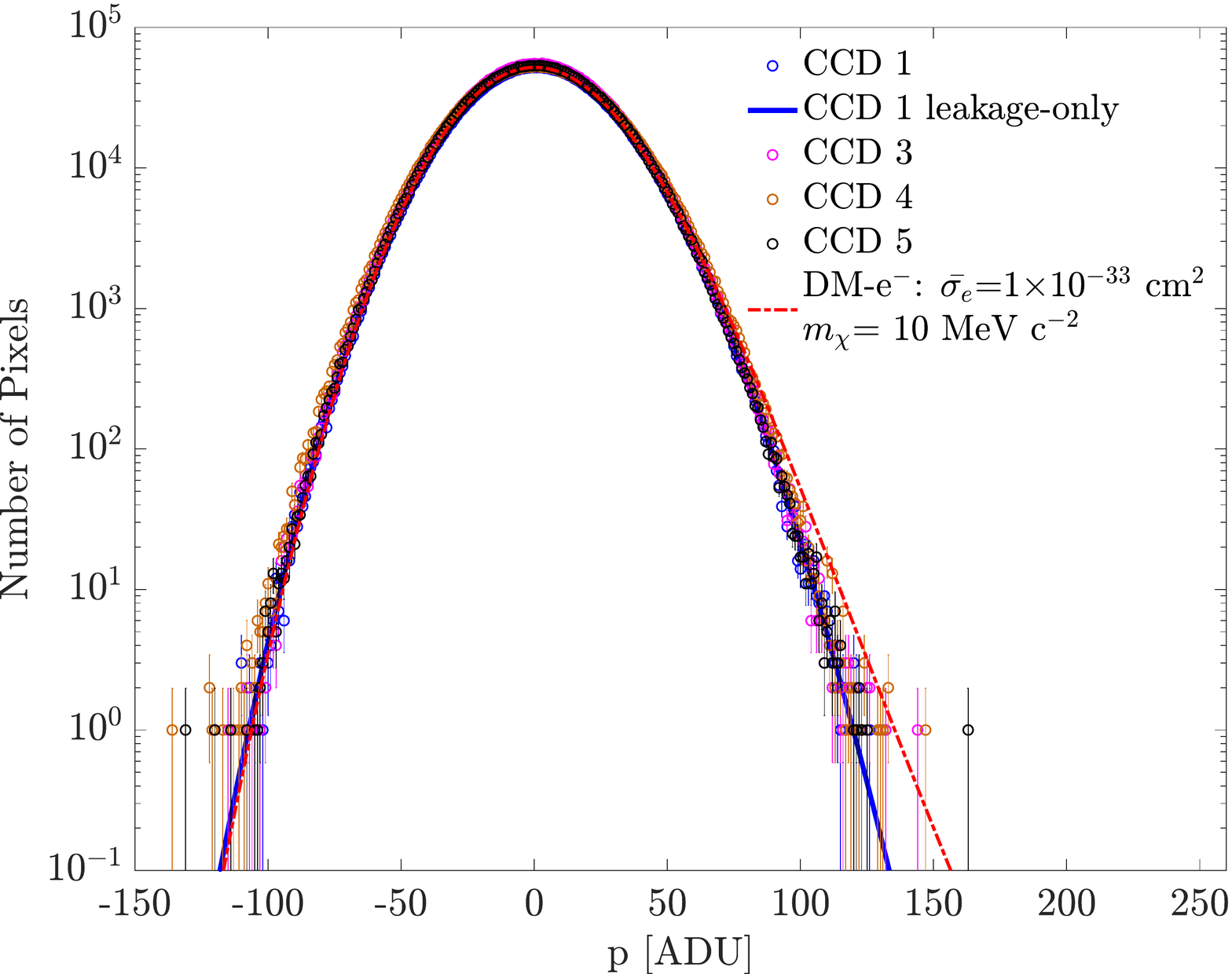}
\includegraphics[width=0.51\linewidth]{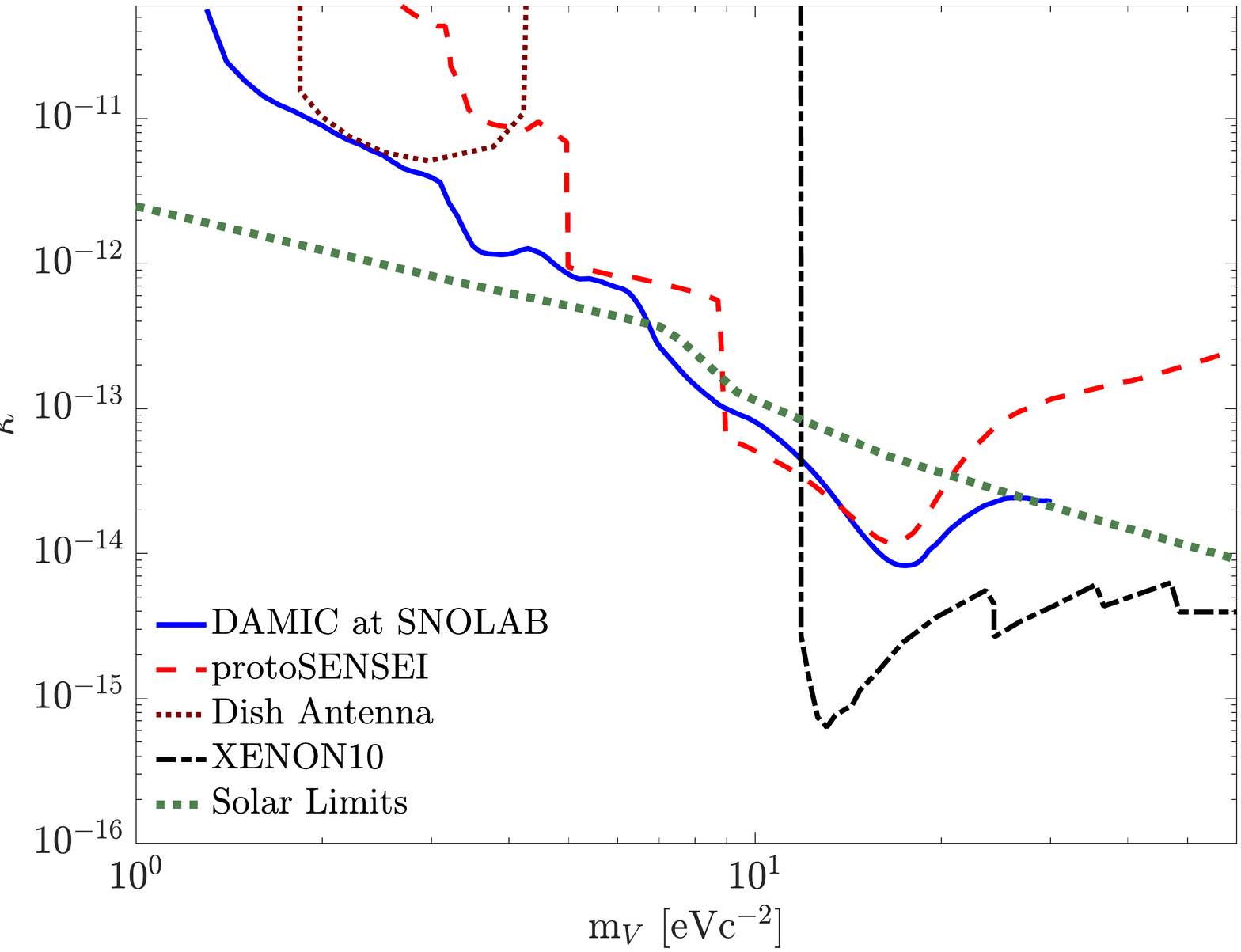}
\caption[]{Left: Distribution of pixel signal in four CCDs selected for the analysis in~\cite{DMelectron_scattering}: in blue the best fit result for the leakage-only model (no DM-e−) and in dashed ref an example fit for the leakage-current plus DM-e− model with $\sigma_{e} = 10^{-33}$~cm$^2$,m$_\chi = 10$~ MeVc$^{−2}$ and F$_{DM} = 1$. Right: 90\% C.L. constraints upper limits on the hidden-photon DM kinetic mixing parameter κ as a function of the hidden-photon mass $m_v$.}
\label{fig:hidden}
\end{figure}

\section{Electron scattering and hidden-photon absorption}
\begin{figure}[b!]
\includegraphics[width=\linewidth]{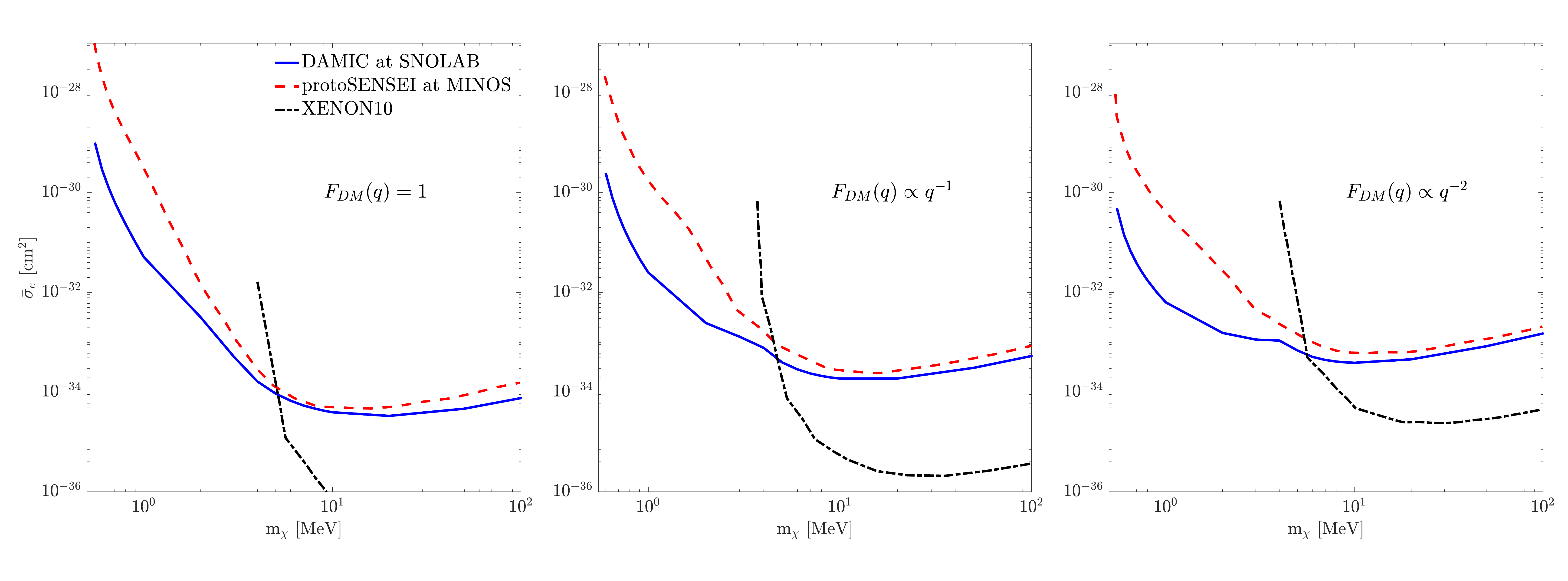}
\caption[]{90\% C.L upper limits on the DM-electron free scattering cross section as a function of DM mass $m_{\chi}$ for F$_{DM} \propto q^{−n}$ for the case n = 0,1,2~\cite{DMelectron-scattering}.}
\label{fig:electronscatt}
\end{figure}
As already mentioned, light DM and hidden sector candidates can be detected through their interaction or absorption with electrons in the target material. This typically allow to  explore region of DM masses down to MeV. 
In both studies the expected signal is observed as a deviation of the pixel signal from the white noise distribution on the positive tail. An example of predicted signals is shown in Fig.~\ref{fig:hidden} where the pixel signal distribution is fit with a gaussian function (blue) and an example of  function describing the hypothesis of white noise plus the signal from a hidden-photon absorption or DM-electron scattering (red).  See the paper~\cite{DMelectron-scattering} for more details on the assumed models shown in figure. 
  
A first study for the hidden photon search was performed in~\cite{HiddenPhotons} with one gram / day exposure. The updated analysis uses 38 images taken specifically acquired with a long exposure (10$^5$~s per image) since 2017 for a  total exposure of 200~g~day. 
The processing of the image and data selection to exclude possible instrumental effects or real detected signals from ordinary particles, are described in details in~\cite{DMelectron-scattering}. 
The upper limits at 90\% C.L. on the hidden-photon kinetic mixing\cite{hidden_mixing}  $k$ as a function of hidden-photon mass $m_v$ are shown in Fig.~\ref{fig:hidden} (right). These are currently the most stringent bounds for masses $m_{v} < 10$~eV. 
A similar approach is followed for DM-electron scattering analysis using 38 exposures collected in 2017, each one as long as $10^5$~s. 
The model is based on~\cite{essig} and the search is performed for candidates with masses (m$_\chi$)  between 0.5 and 100 MeV. Three different values of form factor (F$_{DM}$) have been used to model the transfer momentum ($q$) in the interaction F$_{DM} \propto q^{n}$ with n=0,1,2. The case  n = 0 corresponds  to point-like interactions with heavy mediators or a magnetic dipole coupling,  n = 1  to an electric dipole coupling, and n = 2 to massless or ultra-light mediators~\cite{essig}. 
The 90\% C.L. constraints on the DM-electron cross section are shown in Fig.~\ref{fig:electronscatt} compared to the best results from other direct-detection experiments results: protoSENSEI at MINOS (dashed line)\cite{protoSENSEI,essig2} and XENON10 data (dashed-dotted line)\cite{Xenon10}). The DAMIC limits restrict further the region of DM masses between 0.6 MeV and 6 MeV. 

\begin{figure}[tbh]
\centering
\includegraphics[width=0.9\linewidth]{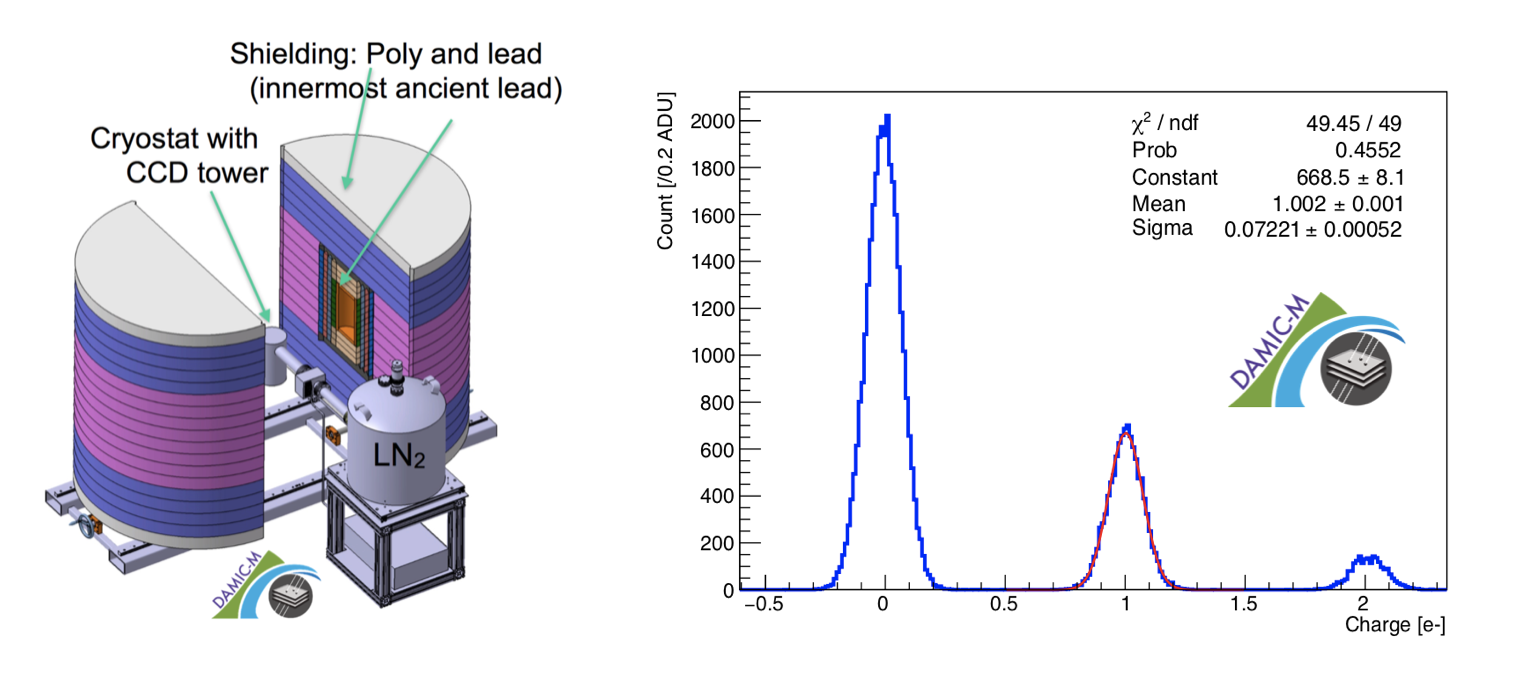}
\caption[]{Left: Sketch of the DAMIC-M design.  Right: single electron resolution measured on a prototype of DAMIC-M skipper-CCDs  (with 1000$\times$6000 pixels and 675$\rm{\mu m}$ thickness). A resolution of 0.07 electrons is achieved. }
\label{fig:damicm}
\end{figure}
\section{Future plans: DAMIC at Modane}
The next step in the DAMIC program is the construction of DAMIC-M at the Modane underground laboratory (LSM)  in France. The new detector will improve the sensitivity to low-mass DM by:  (1) innovating the detector technology to achieve sub-electron resolution on the signal measurements, (2) using a kg-size silicon target  and (3) reducing the background level to a fraction of dru. 
The new CCD technology -- namely ``skipper-CCD'' -- has been designed by Lawrence Berkeley National Laboratory and successfully tested in the SENSEI project~\cite{SENSEI}. The crucial innovation in these devices is the non-destructive, repetitive measurement of the pixel charge, which results in the high-resolution detection of a single electron and unprecedented sensitivity to light DM, because only few eV energies are enough to free an electron in silicon. By counting individual charges in a detector with extremely low leakage current, DAMIC-M will enhance by several orders of magnitude the sensitivity for the hidden sector exploration. A high-resolution detection of a single electron (measured as $\sigma = 0.07$~e$^-$ in Fig.~\ref{fig:damicm}, right) has been achieved with DAMIC-M thick CCDs. 
This resolution and the low leakage current will allow us to reach a threshold down to 2 or 3 electrons ($\sim$3~eV$_{ee}$).

The main challenge is the reduction of the background to a fraction of dru. The background model obtained for DAMIC at SNOLAB, the subsequent measurements of CCD samples and the investigation of the CCD fabrication process provide guidance how this can be achieved. We will screen all materials used in the detector, implement proper handling and cleaning procedures and use the CCD unique spatial resolution to identify background signals.  The main contributions to the current background are due to Pb-210 deposits in the detector surfaces, the hydrogen capture in the dead layer, the cosmogenic activation in the silicon bulk and the radioactive contamination in the copper and in the kapton cables. To minimize cosmogenic activation of tritium inside the silicon material, we will strictly limit the exposure to cosmic ray neutrons by avoiding air transport, storing material either underground or in a shielded container. 
Electro-formed copper will be used  for the CCD frames and for the cryostat with a significant lower content of impurities. Moreover, R\&D activities are on going to improve the material selection for kapton cables and ultra-low-background pico-coaxial cables are under investigation. 

The DAMIC-M project has started in 2018. The detector design and new readout electronics are being developed~\cite{DAMICM-TAUP2019}. The detector will be assembled, tested and operated in a radon-free air, clean room environment at LSM and commissioned in the second half of 2023. 
A prototype using thick skipper-CCDs for DAMIC will be installed at LSM during 2020, with the aim of testing the performance of DAMIC-M skipper-CCDs and perform background measurements at LSM. 

\begin{figure}[tbh]
\centering
\includegraphics[width=0.49\linewidth]{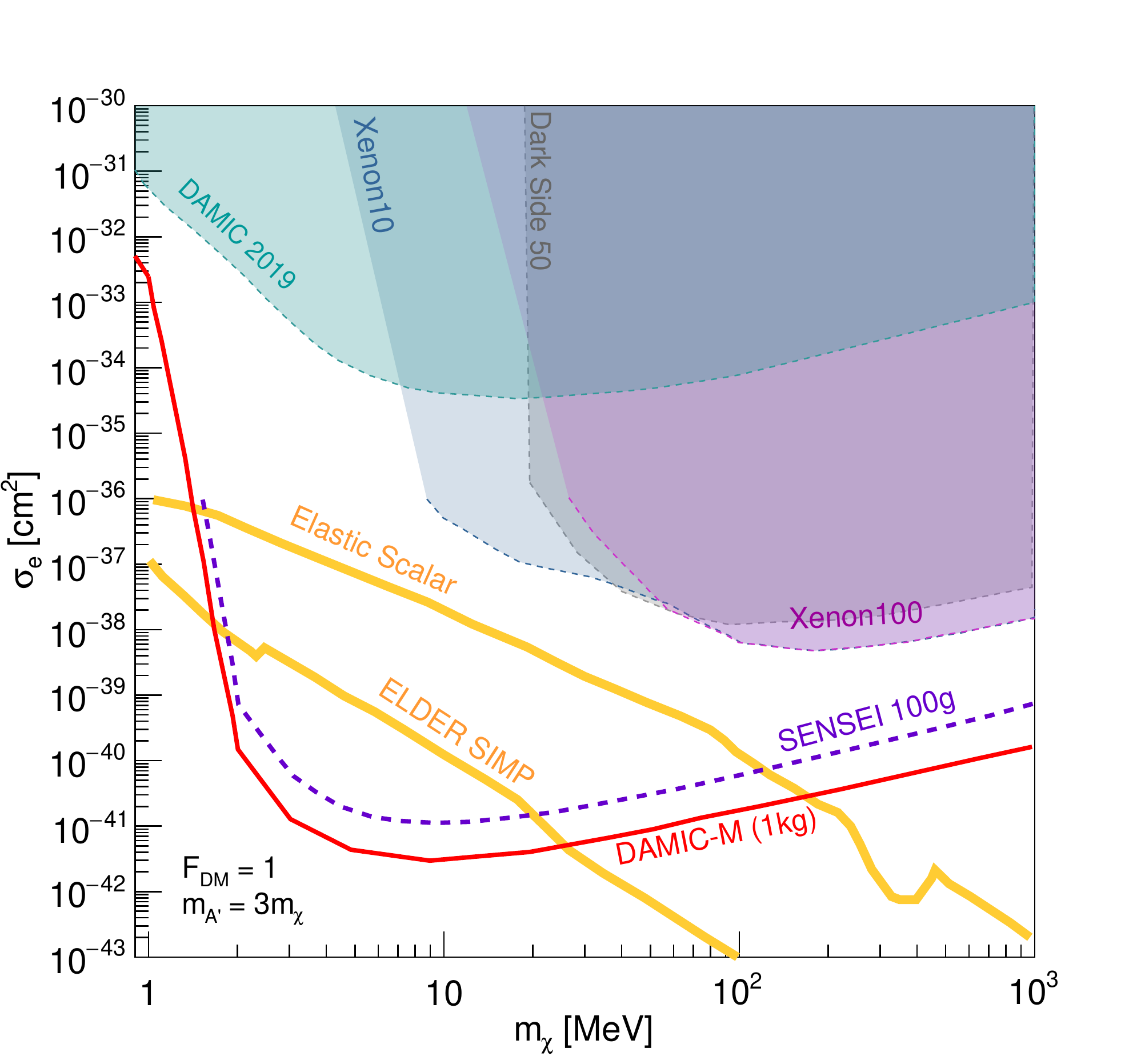}\hfill
\includegraphics[width=0.49\linewidth]{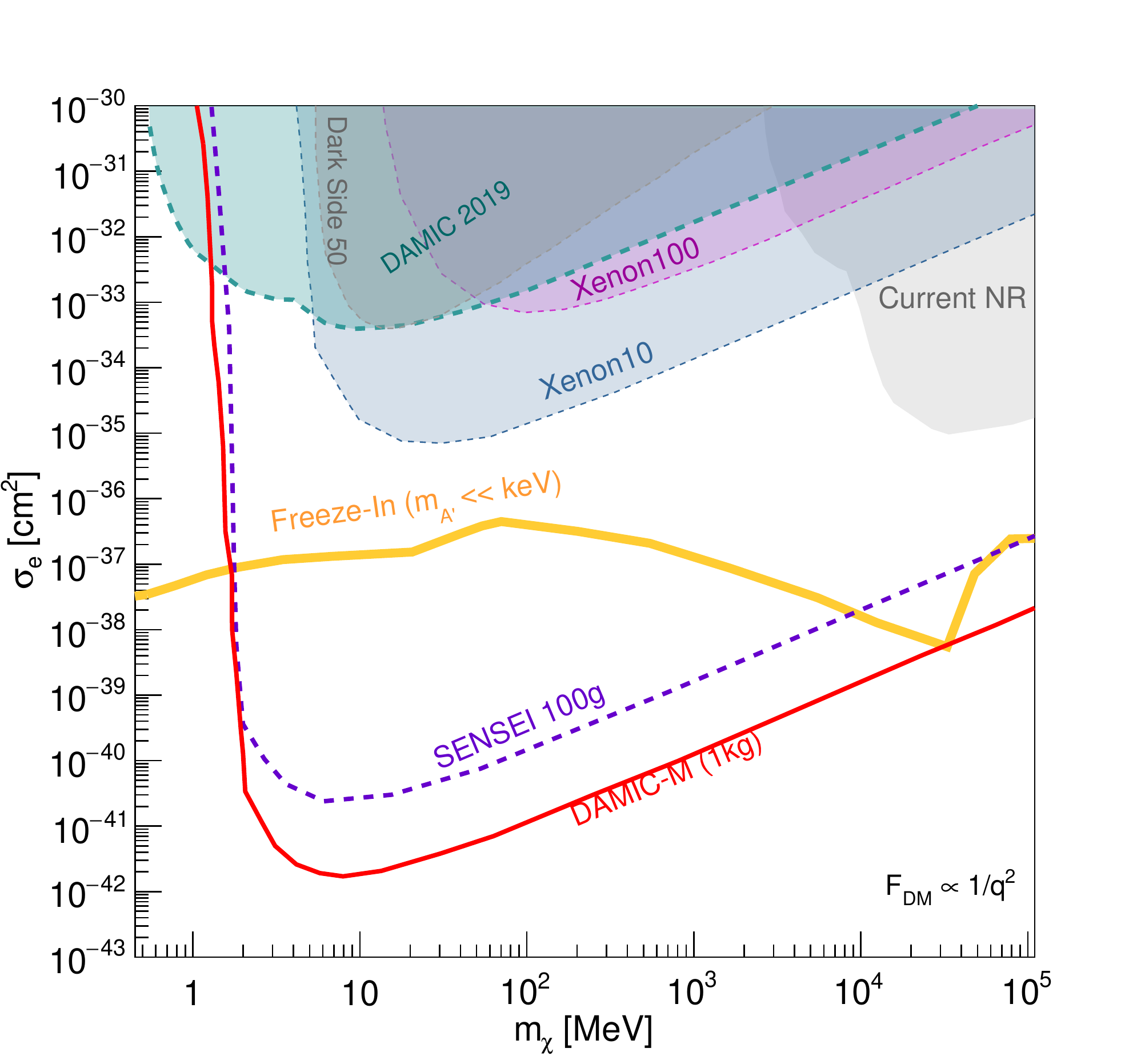}
\caption[]{DM scatters through a heavy (left) and light (right) mediator A′. Theoretical models are shown as thick yellow lines. From~\cite{science}.}
\label{fig:results}
\end{figure}

Assuming that the conditions on the background and the sub-electron resolution are met, DAMIC-M is expected to make a leap forward in the detection of sub-GeV DM particles. The sensitivity for WIMPs-nucleon cross section is shown in Fig.~\ref{fig:results} (right, solid line) with 1~kg year exposure. 
The potentiality of DAMIC-M will be fully exploit in the DM-electron scattering: the lower energy threshold will allow us to probe a large region of parameter space for dark matter particle in the hidden sector with masses between 1~MeV/c$^2$ and 1~GeV/c$^2$. Fig.~\ref{fig:results} (taken from Ref.~\cite{science}) shows the expected sensitivity in the case of a light (right) and heavy (right) mediator (A') in the dark sector. Theoretical models are shown as thick yellow lines.   Current excluded region are dashed and expected results from future experiments are shown as dashed and pointed lines depending on their time-scale. 

\end{document}